\documentclass[aps,pre,twocolumn,groupedaddress,showpacs]{revtex4-1}
\usepackage{bm,graphicx,epstopdf}
\begin{document}

\title{Periodic structures in binary mixtures enforced by Janus particles}

\author{Alexei Krekhov}
\email[]{alexei.krekhov@uni-bayreuth.de}
\author{Vanessa Weith}
\author{Walter Zimmermann}
\affiliation{Physikalisches Institut, Universit\"at Bayreuth, 95440 Bayreuth, Germany}

\date{\today}

%
%
\begin{abstract}
Phase separation in binary mixtures in the presence of Janus particles has been studied in terms of a Cahn-Hilliard model coupled to the Langevin equations describing the particle dynamics.
We demonstrate that the phase separation process is arrested leading to unexpected regular stripe patterns in the concentration field.
The underlying pattern forming mechanism has been elucidated: The twofold absorption properties on the surface of Janus particles with respect to the two components of a binary mixture trigger in their neighborhood spatial concentration variations.
They result in an effective interaction between the particles mediated by the binary mixture.
Our findings open a route to design composite materials with nanoscale lamellar morphologies where the pattern wavelength can be tuned by changing the wetting properties of the Janus particles.
\end{abstract}


\maketitle

\begin{figure}
\centering
$t=10$\hspace*{3.0cm}$t=30$\\
(a)\includegraphics[width=3.5cm]{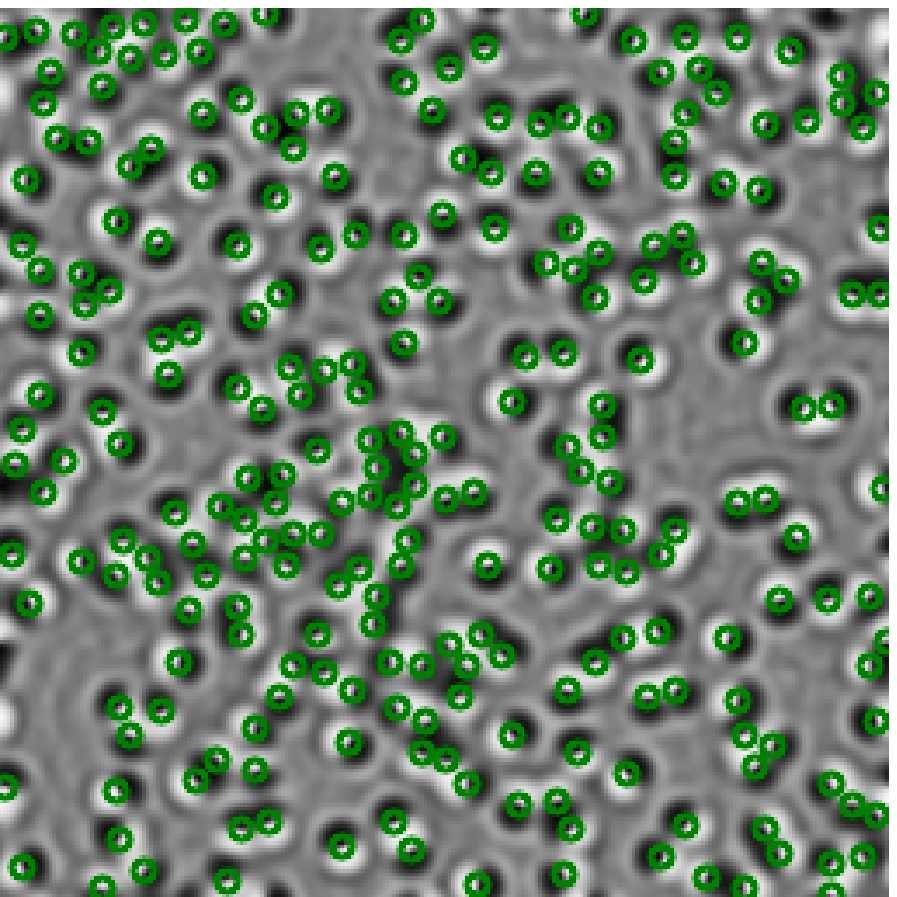}
(b)\includegraphics[width=3.5cm]{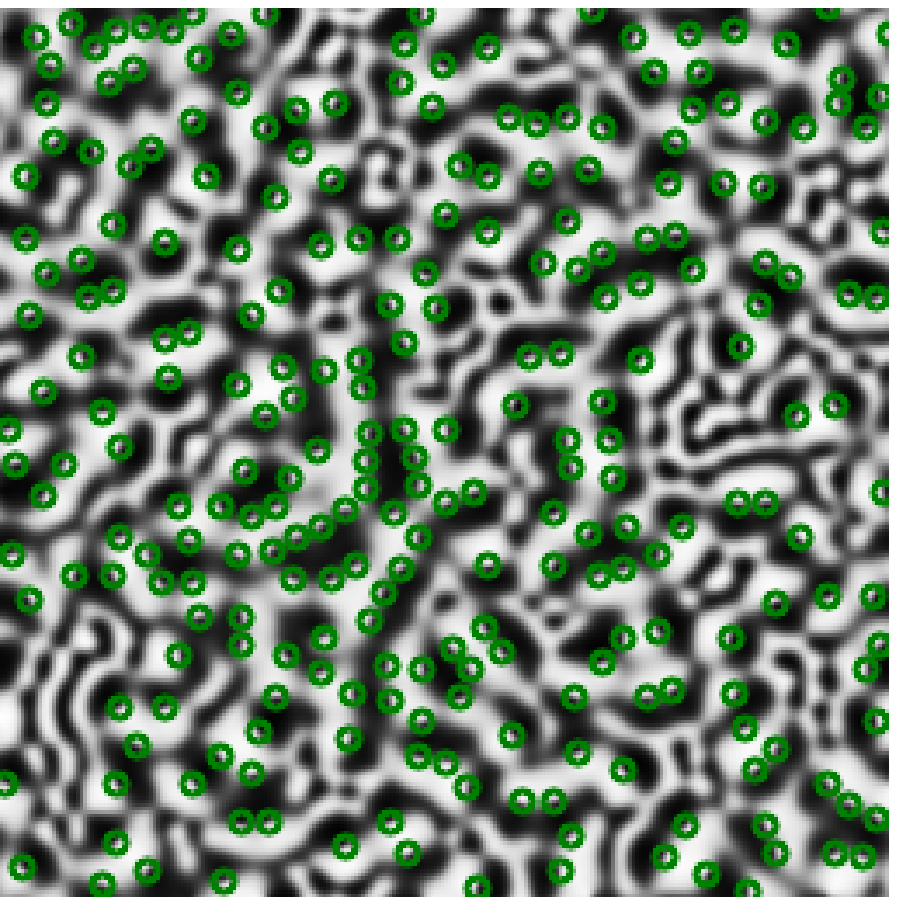}\\
$t=10^3$\hspace*{3.0cm}$t=10^5$\\
(c)\includegraphics[width=3.5cm]{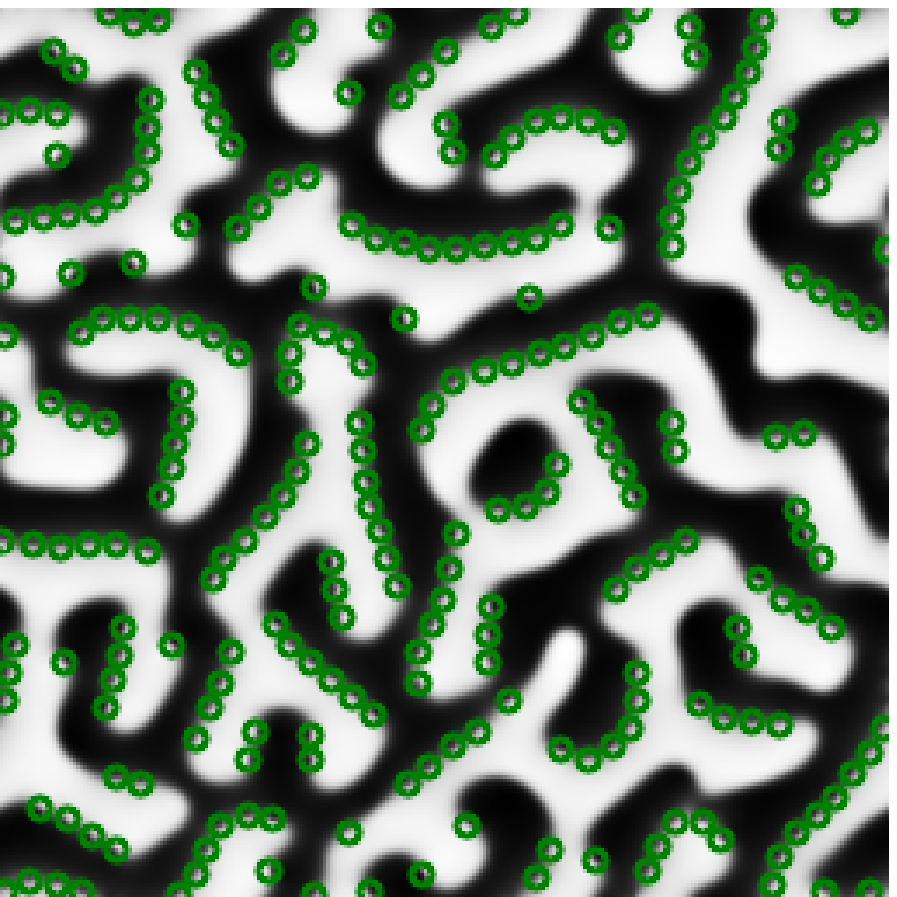}
(d)\includegraphics[width=3.5cm]{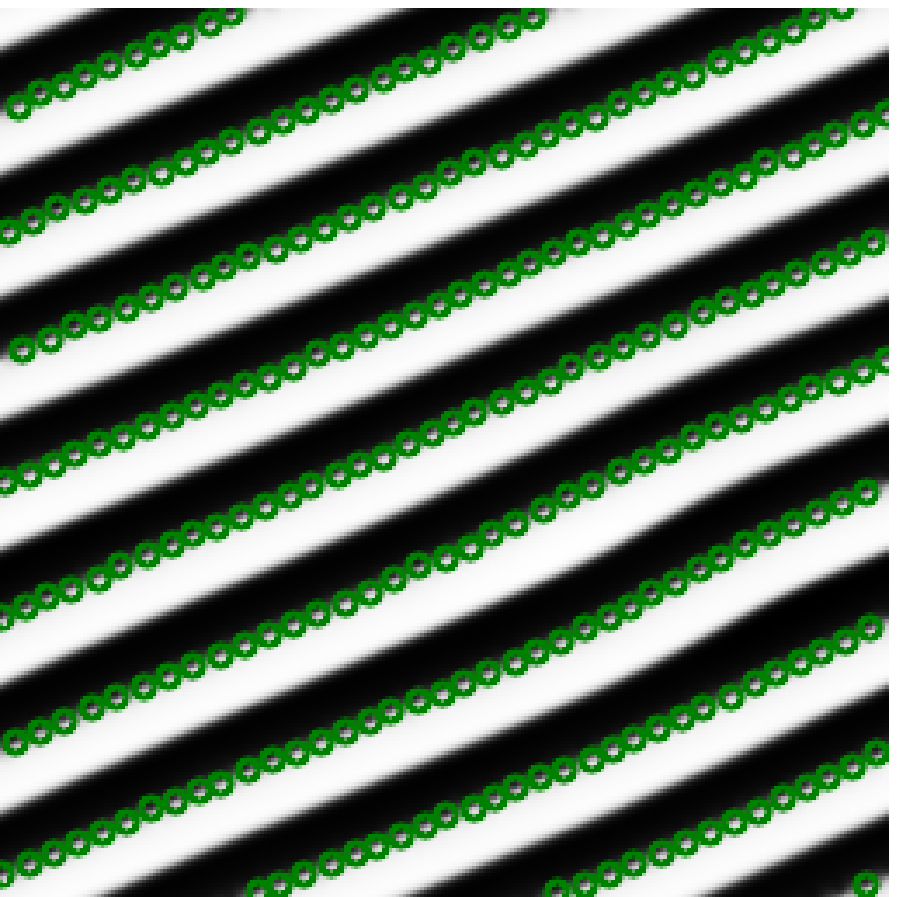}\\
\vspace*{-0.2cm}
\caption{\label{fig:n256_janus}
Snapshots of phase separation in a binary mixture in the presence of Janus particles at times $t=10$ (a), $30$ (b), $10^3$ (c), and $10^5$ (d) after the quench.
$A$- and $B$-rich domains are represented by bright and dark regions and the Janus particles by circles.
Side length $L =256$; number of particles $N =256$.
}
\end{figure}

%
%
Pattern formation in multiphase systems via phase separation is an important interdisciplinary research topic including studies of a large variety of systems, such as alloys, inorganic glasses, polymer blends, biological membranes, bacterial systems, etc.
While for a large number of pattern forming systems in hydrodynamics or with chemical reactions an intrinsic spatial wavelength is characteristic \cite{Cross:1993}, phase separation leads commonly to disordered patterns where the average size of domains with uniform concentration grows in time \cite{Bray:1994}.
The search for mechanisms to gain control over the disorder and the length scale in the phase separation is a central current topic.
In particular in technological applications, ranging from polymer electronics \cite{Sirringhaus:2005, Fichet:2004} to bioactive patterns \cite{Voros:2005}, a control of the arrangement and the size of domains which form the functional elements is crucial for the device performance.
Recently, it has been demonstrated that particles with isotropic wetting properties added to binary mixtures have a strong influence on the phase separation \cite{Chung:2005, Herzig:2007}.
When the particles are equally wetted by the two components they are sequestered at the interfaces, which leads to a structural arrest of phase separation and to the formation of bijels \cite{Stratford:2005, Cates:2008}.
However, the related structures are highly disordered in space.
Here we propose a promising strategy to create periodic structures in binary mixtures by adding Janus particles at low concentrations.
One half of a Janus particle surface is assumed to be preferentially wetted by the $A$- and the other one by the $B$ component of the mixture.
An optimal adjustment of the anisotropic wetting properties of Janus particles to a specific binary mixture is feasible nowadays due to an unprecedental development in particle synthesis \cite{Perro:2005, Walther:2008.1}.
Our analysis is based on the well established combination of a Cahn-Hilliard model with Langevin equations describing the particle dynamics \cite{Ginzburg:1999.1, Balazs:2000.1}.
Typical snapshots of an ``numerical experiment'' in 2D are presented in Fig.~\ref{fig:n256_janus}, where we have found indeed at late stage regular stripe patterns.
The Janus particles accumulated at every second $A|B$ interface and assembled in straight chains.
Further coarsening is then prevented and the patterns have practically reached the equilibrium state.
%

%
%
%

%
%
The free energy of the system, $F= F_{m} + F_{p p} + F_{m p}$, consists of three contributions.
Here $F_{m}$ is the free energy of an incompressible binary mixture, $F_{p p}$ describes the particle-particle interactions, and $F_{m p}$ the mixture-particle interaction.
$F_{m}$ is given by the standard Ginzburg-Landau functional
\begin{eqnarray}
\label{eq:F_m}
F_{m} = \int d{\bf r} \left[
-\frac{1}{2} \varepsilon \psi^2 + \frac{1}{4} u \psi^4 + \frac{1}{2} K (\nabla\psi)^2 
\right] \,,
\end{eqnarray}
where $\psi$ is a real conserved order parameter proportional to the local concentration difference ($\psi >0$ in the $A$-rich and $\psi <0$ in the $B$-rich phase).
The control parameter $\varepsilon$ measures the relative distance to the critical temperature.
Phase separation takes place for $\varepsilon >0$, while for $\varepsilon <0$ the mixture remains homogeneous ($\psi =0$).
The free energy $F_{p p}$ corresponds to the particle-particle repulsive interactions,
\begin{eqnarray}
\label{eq:F_pp}
F_{p p} = \sum \limits_{i} \sum \limits_{j\ne i} U({\bf R}_i-{\bf R}_j) \,,
\end{eqnarray}
where ${\bf R}_i$ ($i=1 \dots N$) denote the center of mass positions of particles.
For distances $|{\bf R}_i-{\bf R}_j|$ larger than an interaction radius $R_0$ (which represents roughly the particle diameter) the potential $U({\bf R}_i-{\bf R}_j)$ is set to zero; otherwise it has been chosen as follows:
\begin{eqnarray}
\label{eq:U}
U({\bf R}_i-{\bf R}_j) = U_0 \left( R_0 - |{\bf R}_i-{\bf R}_j| \right) / |{\bf R}_i-{\bf R}_j|^2 \,.
\end{eqnarray}
The strength of the repulsion is characterized by $U_0 >0$.
The crucial coupling between the local concentration of the mixture $\psi$ and the wetting properties of the particles is given by the free energy $F_{m p}$:
\begin{eqnarray}
\label{eq:F_mp}
F_{m p} = \int d{\bf r} \sum_i
\int d{\bf s}_i V({\bf r}-{\bf s}_i) [\psi({\bf r}) - \psi_s({\bf n}_i)]^2 \,,
\end{eqnarray}
containing an integral over the surface of the $i$-th particle with the surface element $d{\bf s}_i$.
To distinguish the preferential absorption $\psi_s$ at the two different surface parts of the Janus particle we have introduced a unit vector ${\bf n}_i$; it is perpendicular to the ``equatorial plane'' of the particle and points to the ``hemisphere'' preferentially wetted by the $A$-component ($\psi_s >0$).
The short-range wetting potential $V({\bf r})$ is described by
\begin{eqnarray}
\label{eq:Vpot}
V({\bf r}-{\bf s}_i) = V_0 \exp(-|{\bf r}-{\bf s}_i|/r_0) \,,
\end{eqnarray}
where $V_0 >0$ is a measure of its strength and $r_0$ is a microscopic length scale of the wetting interactions.
To minimize $F_{m p}$ the concentration $\psi$ around a Janus particle tends to match the preferential values $\psi_s({\bf n}_i)$ at its two surface parts.
Thus, in contrast to isotropic particles, Janus particles unambiguously trigger phase separation in their vicinity as indicated in Fig.~\ref{fig:n256_janus}(a).
The dynamics of phase separation is described by a generalized Cahn-Hilliard model,
\begin{eqnarray}
\label{eq:CH}
\frac{\partial \psi}{\partial t} = 
M \nabla^2 \frac{\delta }{\delta \psi} (F_m + F_{m p}) \,,
\end{eqnarray}
where $M$ is the ``mobility'' of the component $A$ with respect to $B$.
The translational and rotational Langevin dynamics of the Janus particles is coupled to the local concentration
\begin{eqnarray}
\label{eq:part_r}
&& \frac{\partial {\bf R}_i}{\partial t} = 
-M_r \frac{\partial}{\partial {\bf R}_i} (F_{p p} + F_{m p}) + {\bm \zeta}_i \,,
\\
\label{eq:part_n}
&& \frac{\partial {\bf n}_i}{\partial t} = 
-M_{\theta} \frac{\partial}{\partial {\bf n}_i} F_{m p} + {\bm \xi}_i \,,
\end{eqnarray}
where $M_r$, $M_{\theta}$ are the respective mobility coefficients.
The Gaussian noise terms, ${\bm \zeta}_i$ and ${\bm \xi}_i$, satisfy the fluctuation-dissipation relations.
The results obtained in the framework of this model turned out to be rather insensitive to the particular form of the potentials in Eqs.~(\ref{eq:U}), (\ref{eq:Vpot}).
It is convenient to non-dimensionalize Eqs.~(\ref{eq:CH})-(\ref{eq:part_n}) by measuring $\psi$ in units of $1/\sqrt{u}$ and lengths and time in units of $\sqrt{K}$ and $K/M$, respectively.
The spinodal decomposition wavelength, $\lambda_s = 2\pi/\sqrt{\varepsilon/2}$, which belongs to the maximal unstable linear mode of Eq.~(\ref{eq:CH}), is a universal characteristic length in the binary mixture \cite{Bray:1994}.
It gives the scale over which the decrease of the free energy due to the concentration fluctuations is balanced by the interfacial energy given by the square gradient term in Eq.~(\ref{eq:F_m}).
We have chosen $\varepsilon =1$ in our simulations such that $\lambda_s =8.88$ and $| \psi | \le 1$.
The twofold wetting behavior of the Janus particles is described as $\psi_s =\pm 1$ at the two surface parts.
For the remaining  parameters we used:
$U_0 =10$, $R_0 =4 (\approx \lambda_s/2)$, $V_0 =1$, $r_0 =2 (\approx \lambda_s/4)$, and $M_r=M_{\theta}=1$.
The size of the particles is considered to be smaller than the spinodal decomposition wavelength which is in particular relevant for the case of Janus nanoparticles dispersed in polymer blends \cite{Walther:2008.1}.
As an example, Janus nanoparticles made via self-assembled block copolymers have a size of $25-50$~nm with the surface interaction range of the same order; the spinodal decomposition wavelength is about $100-200$~nm in typical polymer blends under deep quench (see, e.g., Ref.~\cite{Walther:2008.1}).
Finally moderate noise strengths $0.1$ for ${\bm \zeta}_i$ and ${\bm \xi}_i$ have been used.
For larger values the particle dynamics is dominated by thermal fluctuations leading to disordered structures in distinct contrast to Fig.~\ref{fig:n256_janus}(d).
Numerical simulations of Eqs.~(\ref{eq:CH})-(\ref{eq:part_n}) have been performed in two dimensions on a square with a side length $L =256 (\approx 29 \lambda_s)$ and periodic boundary conditions.
In space we used a finite difference discretization scheme with $\delta L =0.5$ and direct forward time integration with the time step $\delta t =10^{-3}$.
%

\begin{figure}
\centering
(a) \includegraphics[width=8.0cm]{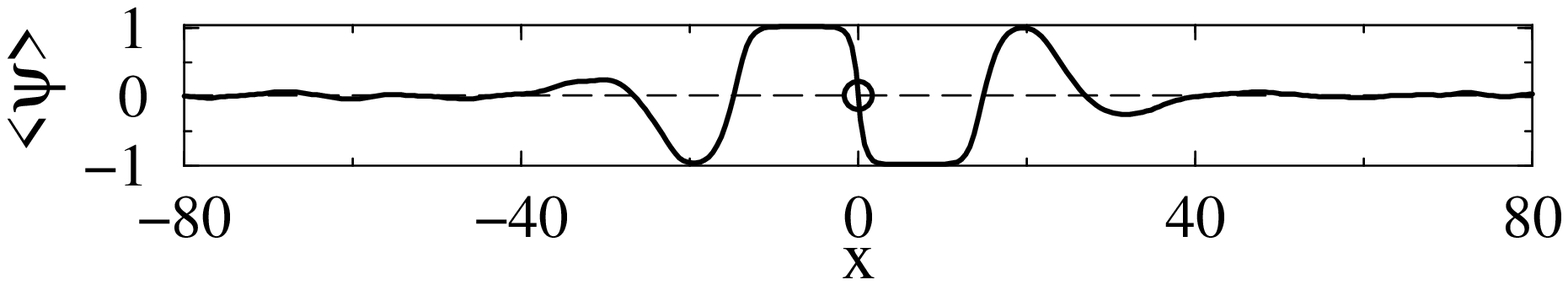}\\
(b) \includegraphics[width=8.0cm]{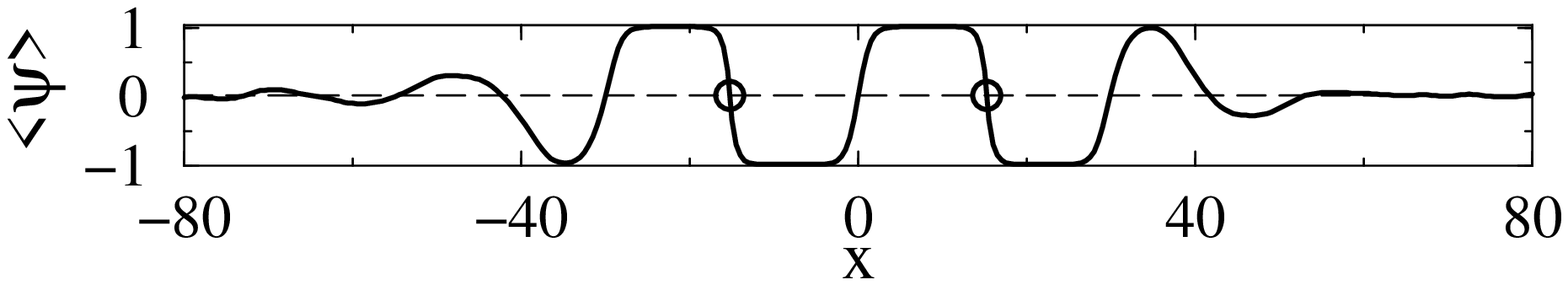}\\
(c) \includegraphics[width=8.0cm]{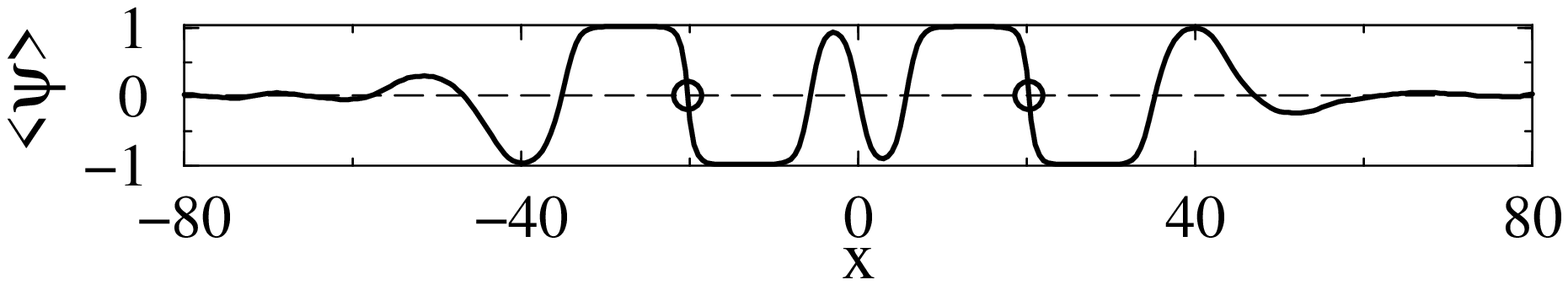}
\caption{\label{fig:avr_psi_np2_l1_rm1_l1_rm1}
Average concentration profiles $\langle \psi \rangle(x)$ near immobile Janus particles at the time $t=10^{4}$ after the quench.
One $(A|B)$ particle at $x =0$ in (a).
Two particles with the same orientation ($A|B$) at the distance $|X_1-X_2| =\Delta X =30$ in (b) and $\Delta X =40$ in (c).
The profiles are the statistical average over $400$ runs with randomly chosen initial conditions for $\psi$.
}
\end{figure}

%
%
Since only in rare cases the particles will experience the soft-core repulsion via Eq.~(\ref{eq:U}), the ordering of Janus particles seen in Fig.~\ref{fig:n256_janus}(d) is caused by their indirect interactions mediated by the concentration field $\psi$.
The mechanism has been elucidated in 1D simulations of Eq.~(\ref{eq:CH}) for $\psi =\psi(x,t)$ with a system size $L =512 (\approx 58 \lambda_s)$.
The time evolution of the particle positions $X_i(t)$ is described by Eq.~(\ref{eq:part_r}), while their  initial ``orientation'' kept fixed.
We use the notion ($B|A$) for a particle with  ${\bf n} = \hat{\bf x}$ and  ($A|B$) for the opposite case, respectively.
To extract an ``effective'' particle-particle interaction beyond the stochastic features of the phase separation process we have analyzed  the statistical average $\langle \psi \rangle(x,t)$ of the field $\psi$.
For that purpose a large number of simulations have been performed where at $t =0$ the initial conditions for $\psi$ have been chosen randomly (noise of the strength $10^{-2}$), while keeping the positions of the particles fixed during the runs.
Averaging over all runs yields then $\langle \psi \rangle(x,t)$.
In the absence of particles or far away from them one has $\langle \psi \rangle =0$.
In contrast, at the position of an immobile Janus particle a steep gradient of $\langle \psi \rangle$ develops, accompanied by spatial concentration oscillations along $x$-axis before eventually $\langle \psi \rangle \to 0$ for large $|x|$, as shown in Fig.~\ref{fig:avr_psi_np2_l1_rm1_l1_rm1}(a).
The wavelength, $\lambda_p$, of the spatial oscillations at late stage is about $30 (\approx 3.4 \lambda_s)$.
In fact the variational derivative $\delta  F_{m p}/\delta \psi$ in Eq.~(\ref{eq:CH}) is obviously responsible for the profile $\langle \psi \rangle$ near the particle.
The derivative leads to an effectively reduced local control parameter $\varepsilon_{eff}(x)  = \varepsilon - 2 V(x-X_i)$ and to an additional forcing term $f(x) = -2 V(x-X_i)$ which are both confined to a small region of the order $r_0$ on both sides of the particle.
%

%
%
\begin{figure}
\centering
\includegraphics[width=8.0cm]{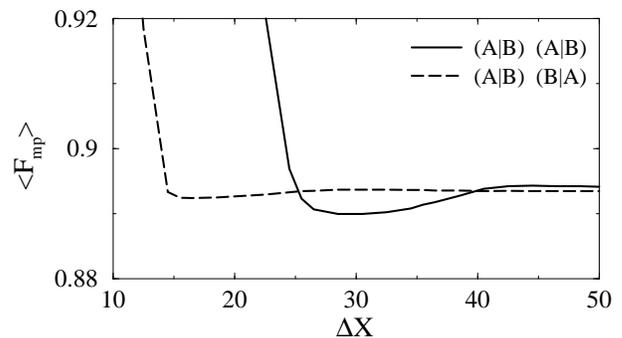}
\caption{\label{fig:avr_Fmp_dx_t1d4}
Average mixture-particle energy $\langle F_{mp} \rangle$ as function of the distance $\Delta X$ between two Janus particles with the same (solid line) and opposite (dashed line) orientations at the time $t=10^{4}$ after the quench.
}
\end{figure}

It is expected that introducing in Fig.~\ref{fig:avr_psi_np2_l1_rm1_l1_rm1}(a) additional particles with the same orientation at the second zero crossings of $\langle \psi \rangle$ ($x \approx \pm \lambda_p$) will lead to a stable configuration since then the total free energy remains essentially constant.
This qualitative argument has been confirmed by an analysis of the average profiles for two immobile particles at positions $X_{1, 2} =\mp \Delta X/2$ [Fig.~\ref{fig:avr_psi_np2_l1_rm1_l1_rm1}(b), (c)].
The averaged mixture-particle energy $\langle F_{mp} \rangle$ is shown in Fig.~\ref{fig:avr_Fmp_dx_t1d4}.
For two Janus particles with the same orientation $\langle F_{mp} \rangle$ has a pronounced minimum at $\Delta X = \lambda_p \approx 30 (\approx 3.4 \lambda_s)$.
The effective force $-\partial \langle F_{m p} \rangle / \partial \Delta X$ between the particles is repulsive for $\Delta X < \lambda_p$ and attractive for $\Delta X > \lambda_p$.
Consequently, the characteristic length $\lambda_p$ may be alternatively interpreted as an equilibrium distance between two particles with the same orientation.
In contrast, for two Janus particles of {\em opposite} orientations $\langle F_{mp} \rangle$ demonstrate very shallow minimum at $\Delta X \approx \lambda_p/2$.
Thus particles only repel each other at $\Delta X < \lambda_p/2$ while they almost do not feel any attractive force at larger distances.
%

%
%
The equilibrium distance, $\lambda_p$, between two Janus particles with the same orientation serves as the key mechanism for their assembling.
Since $\lambda_p$ is much larger than the minimal particle distance $R_0$ [Eq.~(\ref{eq:U})] and the range of the wetting potential $r_0$ [Eq.~(\ref{eq:Vpot})], it is the concentration profile triggered by the Janus particles which causes the effective particle-particle interaction.
The equilibrium distance $\lambda_p$ is determined by the mixture-particle interaction alone.
For $r_0 \to 0$ one has $\lambda_p \to \lambda_s$; $\lambda_p$ grows almost linearly (e.g., with the slope $\approx 1.2 \lambda_s$ for $V_0 =1$ chosen in this paper) when increasing $r_0$.
Note that according to its physical meaning $r_0$ should remain smaller than $\lambda_s$.
Changing the temperature would change the length scale determined by $\lambda_s \propto \varepsilon^{-1/2}$.
Decreasing $V_0$ the amplitude of the $\langle \psi \rangle$ profiles decreases and the effective interaction between Janus particles becomes eventually negligible.
Nevertheless, the particles will still be accumulated at the interfaces as isotropic neutral particles do, where $\psi_s =0$ \cite{Stratford:2005}.
A similar accumulation has been found in a binary mixture with additional surfactant molecules \cite{Kawakatsu:1990}.
Here the coarsening process is arrested when all interfaces are densely occupied by the surfactant molecules.
However, in distinct contrast to our case, the stripe patterns would not develop spontaneously starting from a homogeneous configuration after a quench into the unstable region.
This is obvious since the surfactant molecules do not demonstrate long range interactions which are crucial in our case.
%

%
%
\begin{figure}
\centering
(a)\includegraphics[width=8.3cm]{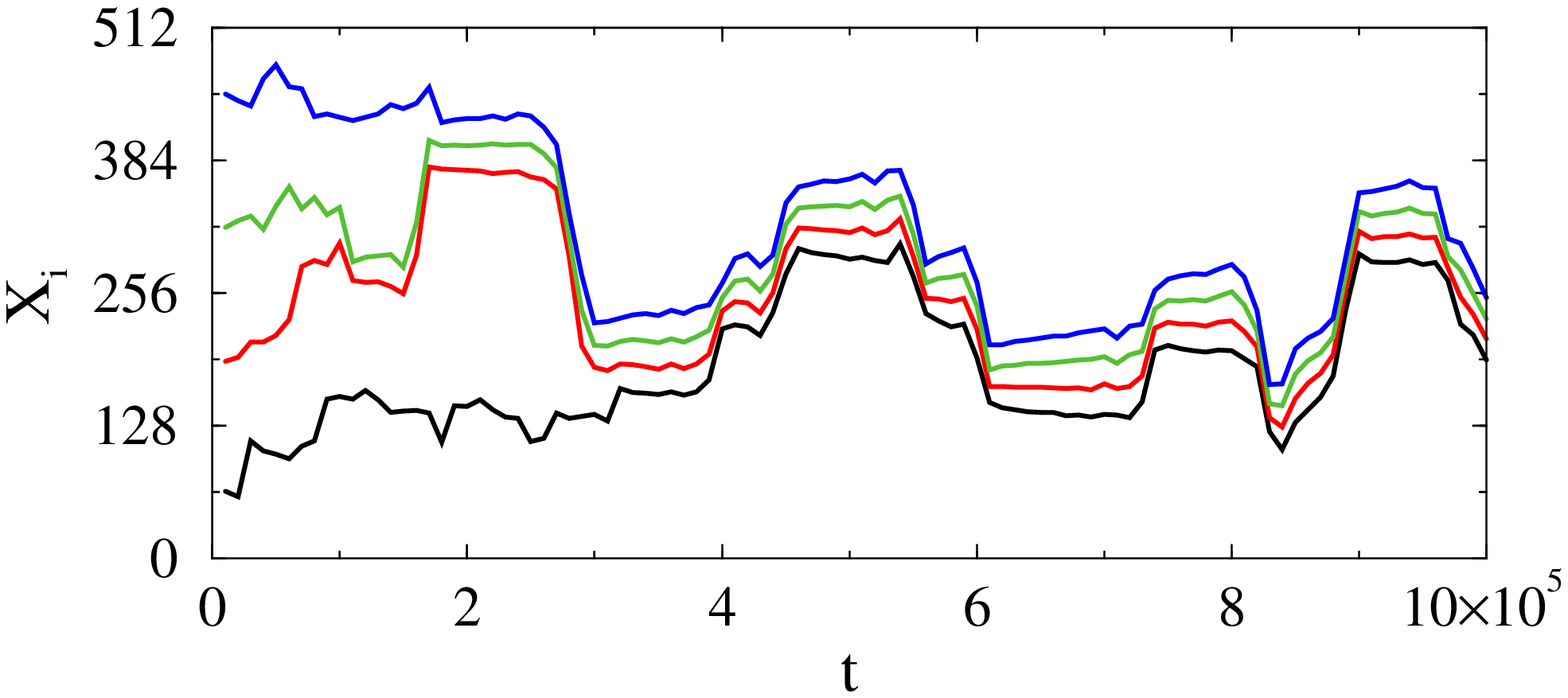}\\
(b)\includegraphics[width=8.0cm]{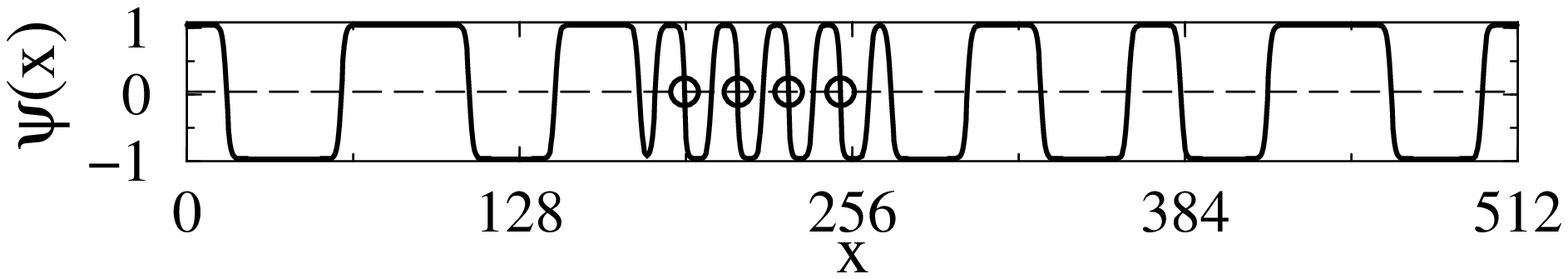}
\caption{\label{fig:pos_t_np4_janus}
(Color online) Trajectories $X_i(t)$ ($i =1 \dots 4$) of four Janus particles with the same orientation located at $t=0$ at distances $\Delta X=128$ in (a).
The concentration profile $\psi(x)$ and the positions of the particles at $t=10^6$ in (b).
}
\end{figure}
To study the combined dynamics between particles and the order parameter $\psi$ we have allowed in a second step the unrestricted movement of the particles.
In Fig.~\ref{fig:pos_t_np4_janus}(a) we present as a typical example the trajectories of four equally oriented Janus particles which were initially placed at an equal distance along the $x$-axis.
When the first two particles have found each other they stay together as time progresses; their distance is determined by $\lambda_p$ as discussed before.
Afterwards the next particle joins this pair also at the equilibrium distance $\lambda_p$ and so on.
Finally all four particles belong to one cluster, which moves as a whole block with the average distance between neighboring particles given by $\lambda_p$ [see Fig.~\ref{fig:pos_t_np4_janus}(b)].
The configuration in the cluster makes obviously use of the basic interaction motif of two particles with free interface in between shown in Fig.~\ref{fig:avr_psi_np2_l1_rm1_l1_rm1}(b).
The phase separation process is thus blocked within the cluster and the coarsening continues only in the free space outside the cluster.
Increasing the number of particles, $N$, the region occupied by the regular array of Janus particles grows accordingly until the cluster  spans the whole system length $L$, i.e., for $N = N_c \approx L /\lambda_p$ the coarsening is completely suppressed.
To block coarsening in the case of oppositely oriented Janus particles which do not exert attractive forces on each other one needs $N = 2 N_c$ particles, such that each interface is occupied by them.
Based on the 1D analysis the formation of stripe patterns observed in 2D simulations as shown in Fig.~\ref{fig:n256_janus} finds now a clear explanation.
Starting from random positions and orientations of the particles, each of them triggers phase separation in its vicinity [Fig.~\ref{fig:n256_janus}(a)].
Afterwards disordered A- and B-rich patches with the characteristic length scale $\lambda_s$ cover the entire domain [Fig.~\ref{fig:n256_janus}(b)].
It is important that all particles are located already after a short time on corresponding interfaces with the particle orientations roughly normal to them which corresponds to the energetically favorite configuration.
With increasing time the particle trajectories will be restricted to their associated interfaces that also move during the coarsening process.
If two particles meet on the same interface they exert only the short-range repulsive force described by $F_{p p}$ [Eq.~(\ref{eq:F_pp})].
The mixture-particle energy contribution $F_{m p}$ [Eq.~(\ref{eq:F_mp})] comes potentially into play if two particles belong to a patch consisting of three interfaces such as the $A|B|A|B$ configuration.
Here one particle sits on the ``left'' $A|B$ interface and the other one on the ``right'' $A|B$ interface with an unoccupied interface in between.
If now the orientations of the particles are almost parallel and the particle distance is of the order of the equilibrium distance $\lambda_p$ they form a bound state in analogy to Fig.~\ref{fig:avr_psi_np2_l1_rm1_l1_rm1}(b).
Thus the orientation of the particles and the distance between them will practically not change during the coarsening process.
Obviously the interfaces of the patch remain parallel as well.
If now a second particle pair of the same configuration approaches the first one along the interface, the two single equilibrium configurations will join and thus straighten the interfaces.
The curvature reduction of the interfaces is also favorable to the square gradient term in $F_m$ [Eq.~(\ref{eq:F_m})].
It should be emphasized, however, that in contrast to the 1D case [Fig.~\ref{fig:pos_t_np4_janus}(b)] a suitable concentration of the particles has to be chosen to generate periodic 2D patterns.
As follows from our previous discussion at least every second interface has to be densely occupied by particles at late stage.
This is confirmed by the simulations shown in Fig.~\ref{fig:n256_janus}(d).
If the particles concentration is too low the ``entropy'' due to the orientational and positional degrees of freedom counteracts the mixture-particle energy contribution, such that regular structures will not develop.
To obtain stripe patterns we need of the order of $N_p = L^2 / (R_0 \lambda_p)$ particles in the system, which corresponds to a particle concentration $c = c_p = N_p \pi R_0^2 / (4 L^2) = \pi R_0/(4 \lambda_p)$.
In our case we had $c_p \approx 0.1$ but it can be reduced by increasing the equilibrium distance $\lambda_p$ determined by the wetting properties of the particles.
Since the distance between particles located on the same interface can vary about $R_0$, the regular stripes can be found already for $c \approx c_p/2$ [Fig.~\ref{fig:n256_janus}(d)].
On the other hand we find also stripe patterns for $c > c_p$ up to $c \approx 2 c_p$  where now all interfaces are occupied with particles having the opposite orientations on adjacent interfaces.
%

%
%
In summary, we have demonstrated that adding Janus particles above a certain concentration to a phase-separating binary mixture drives the system into regular structures with interfacially sequestered particles.
The coarsening process is arrested and the wavelength of the patterns is determined by the wetting properties of the particles.
As to be expected we have also found in our simulations that the formation of stripe patterns becomes much more efficient when all Janus particles have the same orientation.
This can be easily realized utilizing magnetic Janus particles \cite{Zhao:2009, Arita:2013} and applying an external magnetic field.
In addition the regular structures might be pre-conditioned by starting with a directional quench or applying a spatio-temporal periodic temperature modulation \cite{Krekhov:2009, Weith:2009}.
Our results suggest a promising approach to create new structures of composite materials.
A complementary aspect of such systems is related to the self-assembly and ordering of the nanoparticles which is now receiving considerable attention \cite{Hong:2006, Sciortino:2009, Boeker:2007}.
A further interesting aspect found in our simulations is a super-diffusional behavior of a single Janus particle during phase separation; the mean square displacement of the particle has varied like $t^{\alpha}$ with $\alpha > 1.5$.
Super-diffusion of particles is an important issue in turbulent flows where similar preferential concentration of particles in high strain regions has been found \cite{Gibert:2012}.
%

%
%
%
We are grateful to W. Pesch for stimulating discussions and for critically reading the manuscript.
Financial support by the Deutsche Forschungsgemeinschaft Grants SFB~840 and FOR~608 is gratefully acknowledged.
%

%
%
%
%
%

\end{document}